\documentclass[nojss]{jss}

\usepackage[english]{babel}
\usepackage[utf8]{inputenc}
\usepackage{amsmath}
\usepackage[draft]{fixme}
\usepackage{graphicx}
\usepackage{hyperref}
\usepackage{amssymb}
\usepackage{color}
\usepackage{thumbpdf}

\author{Mikkel Meyer Andersen\\Aalborg University \And 
       Poul Svante Eriksen\\Aalborg University}
\title{Efficient Forward Simulation of Fisher-Wright Populations with Stochastic Population Size and Neutral Single Step Mutations in Haplotypes}

\Plainauthor{Mikkel Meyer Andersen, Poul Svante Eriksen}
\Plaintitle{Efficient Forward Simulation of Fisher-Wright Populations with Stochastic Population Size and Neutral Single Step Mutations} 
\Shorttitle{Efficient Simulation of Fisher-Wright Populations}

\Abstract{
In both population genetics and forensic genetics it is important to know how haplotypes are distributed in a population. Simulation of population dynamics helps facilitating research on the distribution of haplotypes. In forensic genetics, the haplotypes can for example consist of lineage markers such as short tandem repeat loci on the Y chromosome (Y-STR). A dominating model for describing population dynamics is the simple, yet powerful, Fisher-Wright model. We describe an efficient algorithm for exact forward simulation of exact Fisher-Wright populations (and not approximative such as the coalescent model). The efficiency comes from convenient data structures by changing the traditional view from individuals to haplotypes. The algorithm is implemented in the open-source \proglang{R} package \pkg{fwsim} and is able to simulate very large populations. We focus on a haploid model and assume stochastic population size with flexible growth specification, no selection, a neutral single step mutation process, and self-reproducing individuals. These assumptions make the algorithm ideal for studying lineage markers such as Y-STR.
}
\Keywords{\pkg{fwsim}, \proglang{R}, $k$-d tree, forensic genetics, Y-STR, stepwise mutation model, single step mutation model}
\Plainkeywords{fwsim, R, k-d tree, Y-STR, stepwise mutation model, single step mutation model} 

\Address{
  Aalborg University\\
  Department of Mathematical Sciences\\
  E-mail: \email{mikl@math.aau.dk}\\
  URL: \url{http://people.math.aau.dk/~mikl/}
}

\begin{document}

\section{Introduction}
\nopagebreak
Simulation of population dynamics is an important tool when studying genetic traits. In both population genetics and forensic genetics it is important to know how haplotypes are distributed in a population. In forensic genetics, the haplotypes can for example consist of lineage markers such as short tandem repeat loci on the Y chromosome (Y-STR). Simulation of population dynamics helps facilitating research on the distribution of haplotypes. A dominating model for describing population dynamics is the simple, yet powerful, Fisher-Wright model (or process) \citep{Fisher1922, Fisher1930, Fisher1958, Wright1931, Ewens2004}. In population genetics, the model also forms the basis for coalescent theory \citep{kingman_coalescent_1982, Hudson2001, hein_gene_2005}.

Because the Fisher-Wright model is widely used in population genetics, efficient simulation algorithms and tools are needed. In this paper we describe the model implemented in the \proglang{R} \citep{R} package \pkg{fwsim} \citep{fwsim025}, which provides an efficient tool for simulating certain kinds of Fisher-Wright populations. The simulation scheme described in this paper is exact (from the Fisher-Wright model) and not approximative like the simulation scheme from the coalescent model \citep{kingman_coalescent_1982, Hudson2001, hein_gene_2005}.

\citep{Ewens2004} is a good reference on different models in population genetics as it explains several models and also gives theoretical results.

First some nomenclature must be introduced. Let a locus (loci in plural) be a specific location on the chromosome. The content of a locus is called an allele, which consists of DNA sequences. Here, we assume that the alleles are short tandem repeats (STRs) \citep{Butler2005-2nd} with values in $\mathbb{Z}$ (in genes, an allele could also just be either of two states, $A$ or $B$, say). A haplotype is a ordered collection of alleles at loci that are transmitted together.

We focus on a haploid model, where each individual is a gamete with a haplotype consisting of $r$ loci. Hence, a haplotype can in this context be thought of as a vector in $\mathbb{Z}^r$. It may for example be an Y-STR haplotype. We assume no selection and the individuals are self-reproducing.

First, the traditional Fisher-Wright model without mutations is described in order to introduce the notation and to make it possible to compare it with our model.

Throughout this paper, whenever there is a mutation process, we assume it to be a neutral (in the sense of no selection) single step mutation process with infinitely many possible allelic states. This model was introduced by \citep{OhtaKimura1973} and some mathematical properties were recently discussed in \citep{Caliebe2010}.

\subsection{Fisher-Wright model without mutation} \label{sec:simple-fisher-wright}
\nopagebreak
Traditionally, a simple Fisher-Wright model, for example as formulated by \citep{Ewens2004}, assumes constant population size and no mutations. A Fisher-Wright model is often characterised by a binomial sampling scheme focusing on individuals (or a multinomial sampling scheme focusing on the entire population), such that a new generation of children is sampled by letting each child choose its parent (and thus its haplotype) uniformly at random. 

Because our interest is aimed at the sampling of populations and not at the genealogy, the focus is now changed from individuals to haplotypes, where identical haplotypes are treated similarly, as we are not interested in the genealogical tree itself, but only in the haplotypes and their counts in the resulting population (and possibly in the intermediate populations, too).

Let $N$ be the constant, known population size and $H$ the set of haplotypes. Denote by $n_i(x)$ the number of haplotypes in the $i$'th generation of haplotype $x \in H$ and $z_{i+1}(x)$ the number of children from haplotype $x \in H$ in generation $i+1$. Because there are no mutations, we have that $n_{i+1}(x) = z_{i+1}(x)$.

The simple Fisher-Wright model arises by assuming that $P\left(\{ n_{i+1}(x) \}_{x \in H} \mid \{ n_i(x) \}_{x \in H}\right)$ is given by
\begin{align} \label{eq:simple-fw}
  \{ n_{i+1}(x) \}_{x \in H} \mid \{ n_i(x) \}_{x \in H}
  \sim
  \text{Multinomial} \left (N, \left \{ \frac{n_i(x)}{N} \right \}_{x \in H} \right ).
\end{align}

A property of the multinomial distribution is that
\begin{align*}
  \mathbf{E} \left [n_{i+1}(x) \mid n_i(x) \right ] = n_i(x) 
\end{align*}
as expected.

We note that the process is a Markov chain with $\vert H \vert$ absorbing states, one for each haplotype.

\section{Model}
\nopagebreak
As mentioned in Section~\ref{sec:simple-fisher-wright}, the model is formulated on the basis of haplotypes instead of individuals, because it is much more efficient when we are interested in the resulting population after a number of generations rather than the genealogy.

The notation from Section~\ref{sec:simple-fisher-wright} is adopted, such that $H_i$ is the set of haplotypes in the $i$'th generation ($H_i$ depends on $i$ due to mutations, which will be introduced below), $n_i(x)$ is the number of haplotypes in the $i$'th generation of haplotype $x \in H_i$, and $z_{i+1}(x)$ the number of children from haplotype $x \in H_i$. Now let $N_i = \sum_{x \in H_i} n_i(x)$ be the population size in the $i$'th generation (instead of a constant population size $N$ as in the simple Fisher-Wright model in Equation~\ref{eq:simple-fw}).

Our model is then a specification of how 
\begin{align*}
  \{z_{i+1}(x)\}_{x \in H_i} \mid \{n_{i}(x)\}_{x \in H_i}
\end{align*}
is distributed, that is, how the haplotypes in the next generation are conditionally distributed given the previous generation. 

Two important features of our model is, that it assumes stochastic population size -- which we believe is a more realistic model -- and allows flexible population growth specification. We believe that the Fisher-Wright model that will be introduced below with stochastic population size also incorporating flexible population growth has not yet been defined like we do in the following. First the modelling of the population size and growth will be described. Afterwards the mutational model will be explained.

\subsection{Population size and growth}
\nopagebreak
Let $N_0$ be the known initial population size. Note that in the traditional Fisher-Wright model, this is assumed to be a constant. 

Then we assume that
\begin{align} \label{eq:population-size}
  N_i \mid N_{i-1} \sim \mbox{Poisson}(\alpha_i N_{i-1})
\end{align}
for $\alpha_i > 0$ ($\alpha_i > 1$ gives growth and $0 < \alpha_i < 1$ gives decline). For example, if $\alpha_i = \alpha$ for all $i$, then
\begin{align*}
  \mathbf{E}[N_i] = \alpha^i N_0 ,
\end{align*}
that is exponential population growth. One could also choose 
\begin{align*}
  \alpha_i =
  \begin{cases}
    \beta, & \text{for $i \leq t$,} \\
    \alpha, & \text{else,}
  \end{cases}
\end{align*}
yielding 
\begin{align*}
\mathbf{E}[N_i] = 
  \begin{cases}
    \beta^i N_0, & \text{for $i \leq t$,} \\
    \beta^t \alpha^{i-t} N_0, & \text{else,}
  \end{cases}
\end{align*}
which for example can be used to get exponential growth up to generation $t$ and afterwards an expected constant population size by setting $\alpha = 1$. 

A possibly more realistic example is logistic population growth, which can be obtained by specifying a maximum population size $N_{max}$, $\alpha \geq 1$, and then setting
\begin{align*}
  \alpha_i = \alpha - \frac{(\alpha-1)N_{i-1}}{N_{max}}  
\end{align*}
as the growth rates. A closed form expression for $\mathbf{E}[N_i]$ in this case seems difficult to obtain.

One could alternatively also create a (possibly decreasing) rate $\alpha_i = f(i)$ for some function $f$. Hence, the specification of growth is rather flexible.

\subsection{Number of children}
\nopagebreak
As mentioned previously, the conditional distribution $\{z_{i+1}(x)\}_{x \in H_i} \mid \{x_{i}(x)\}_{x \in H_i}$ must be specified. We assume that the number of children $z_{i+1}(x_0)$ of a certain haplotype $x_0 \in H_i$ is conditionally independent of the number of children of other haplotypes, given the entire previous generation $\{x_{i}(x)\}_{x \in H_i}$. Thus, only the marginal distribution $z_{i+1}(x_0) \mid \{x_{i}(x)\}_{x \in H_i}$ must be specified.

For each haplotype $x_0 \in H_i$ in the $i$'th generation occuring $n_i(x_0)$ times, we then assume that the number of children $z_{i+1}(x_0)$ is distributed independently of other haplotypes as
\begin{align} \label{eq:number-of-children}
   z_{i+1}(x_0) \mid \{ n_i(x) \}_{x \in H_i} \sim \text{Poisson}(\alpha_{i+1} n_i(x_0)) .
\end{align}
As can be seen, $z_{i+1}(x_0)$ actually only depends on $n_i(x_0)$ and not on the number of all the other haplotypes.

It then follows that $N_{i+1} = \sum_{x \in H_i} z_{i+1}(x)$ (the sum of the number of haplotypes in the $(i+1)$'th generation) conditionally on $\{n_i(x)\}_{x \in H_i}$ follows a $\text{Poisson}(\alpha_{i+1} N_{i})$ distribution, and that
\begin{align*}
   z_{i+1}(x_0) \mid \{n_i(x)\}_{x \in H_i}, N_{i+1} \sim \text{Binomial} \left ( N_{i+1}, \frac{n_i(x_0)}{N_{i}} \right ),
\end{align*}
as expected, which is also true for the simple Fisher-Wright model in Equation~\ref{eq:simple-fw}.

\subsection{Mutation model} \label{sec:model-mutation}
\nopagebreak
As mentioned in the introduction, we assume a neutral (in the sense of no selection) single step mutation process on $\mathbb{Z}$. Instead of just one locus we extend it to $r$ loci, where mutations on loci happen independently. We assume per locus and direction mutation rates. Let
\begin{align*}
  Q = \{ -1, 0, 1\}^r = \underbrace{\{ -1, 0, 1\} \times \cdots \times \{ -1, 0, 1\}}_{\text{$r$ factors}} ,
\end{align*}
where $\times$ denotes the Cartesian product, be the lattice of possible mutations. Let
\begin{align} \label{eq:locus-mut-prob}
  p_j(q) =
  \begin{cases}
    \delta_j & q = -1 \\
    1 - \delta_j - \omega_j & q = 0 \\
    \omega_j & q = 1 \\
    0 & \text{else}
  \end{cases}
\end{align}
denote the mutation probabilities for the $j$'th locus and
\begin{align*}
  p(q) = \prod_{j=1}^r p_j(q_j)
\end{align*}
for a mutation configuration $q = (q_1, q_2, \ldots, q_r) \in Q$ from the fact that mutations are assumed to happen independently across loci.

Let
\begin{align*}
  C_{i+1} = \bigcup_{\substack{q \in Q\\x_1 \in H_i}} \{ x_1 + q \}
\end{align*}
be all possible candidate haplotypes for the $(i+1)$'th generation.

Our model with mutations is then
\begin{align} \label{eq:mut-model}
  n_{i+1}(y_0) \mid \{n_i(x)\}_{x \in H_i} \sim \text{Poisson} \left ( \alpha_{i+1} \sum_{q \in Q} p(q) n_i(y_0 - q) \right ) 
  \quad
  \text{for all $y_0 \in C_{i+1}$},
\end{align}
resulting in $N_{i+1} \mid N_i \sim \text{Poisson} ( \alpha_{i+1} N_i )$ as assumed in Equation~\ref{eq:population-size} because 
\begin{align*}
  \sum_{y_0 \in C_{i+1}} \alpha_{i+1} \sum_{q \in Q} p(q) n_i(y_0 - q)
  &= \alpha_{i+1} \sum_{q \in Q} p(q) \sum_{y_0 \in C_{i+1}} n_i(y_0 - q) \\
  &= \alpha_{i+1} \sum_{q \in Q} p(q) N_i \\
  &= \alpha_{i+1} N_i .
\end{align*}

Another way to formulate an equivalent model, which will be used in the implementation, is as follows. Let $m_{i+1}(x, x + q)$ denote the number of mutants mutating from $x$ to $x+q$ in the transition from the $i$'th generation to the $(i+1)$'th generation and 
\begin{align*}
  M_{i+1}(x) = \{ m_{i+1}(x, x + q) \}_{q \in Q}
\end{align*}
the number of mutants for all possible configurations in $Q$.

Then assume that $\{ M_{i+1}(x) \}_{x \in H_i}$ are conditionally independent given $\{ z_{i+1}(x) \}_{x \in H_i}$, thus only the marginal distribution is to be specified. If we model this conditional marginal distribution as
\begin{align} \label{eq:mut-model-alternative}
  M_{i+1}(x_0) \mid \{ z_{i+1}(x) \}_{x \in H_i}
  \sim \text{Multinomial} \left ( z_{i+1}(x_0) , \{ p(q) \}_{q \in Q} \right ),
\end{align}
and set
\begin{align*}
  n_{i+1}(x) = \sum_{q \in Q} m_{i+1}(x - q, x) ,
\end{align*}
we get a model equivalent to the one specified in Equation~\ref{eq:mut-model}.

\subsection{Absorbing state}
\nopagebreak
The model in Equation~\ref{eq:mut-model} (or the equivalent model in Equation~\ref{eq:mut-model-alternative}) has positive probability of dying out, because the Poisson distribution has probability mass in $0$ for every parameter value. This means that population size $0$ is an absorbing state. Also note that this absorbing state is independent of the mutation rate, as the population size is independent of the mutation rate.

\section{Implementation}
\nopagebreak
In this section, some implementation details are discussed. As already mentioned, the described model is implemented in the \proglang{R} \citep{R} package \pkg{fwsim} \citep{fwsim025} using the \proglang{C} programming language. The package \pkg{fwsim} is released under the BSD license. 

First some implementation details are explained and then a few examples are given.

\subsection{Haplotype container} 
\nopagebreak
Each generation consists of a number of haplotypes, each with a count of the number of times it is present in the generation. These haplotypes are saved in a data container. This data container is a so-called $k$-d tree \citep{Bentley1975} (this abbrivation stands for $k$ dimensional tree), which is a generalisation of a binary search tree. Whereas binary search trees are for one dimensional points (numbers), $k$-d trees are for $k$ dimensional points (vectors). Like binary search trees, the time complexity for insertion and searching in a $k$-d tree is $O(\log n)$ for a tree with $n$ nodes.

For each generation, a new $k$-d tree is created and nodes inserted or updated as the haplotypes are evolved one at a time. A node in the tree contains both the point (haplotype) and additional information, which here is only a count (of the number of individuals having this particular haplotype).

The implementation of $k$-d trees is based on \url{http://code.google.com/kdtree} released under the BSD license, but has been heavily modified for example by changing some data structures and adding node searching and updating functionality.

\subsection{Mutation model} \label{sec:implementation-mutation}
\nopagebreak
In this section, the implementation of the mutation model defined in Section~\ref{sec:model-mutation} is described.

The mutation model is implemented by dividing the number of children Equation~\ref{eq:number-of-children} into categories depending on the number of times they mutate. There are $r+1$ categories, namely for $d = 0, 1, \ldots, r$ mutations on the $r$ loci. Because this is the stepwise mutation model, only one mutation can happen per locus at a time.

As before, $z_{i+1}(x)$ is the number of children from haplotype $x \in H$. Let $z_{i+1}^d(x)$ be the number of children in the $d$'th category such that $z_{i+1}(x) = \sum_{d=0}^r z_{i+1}^d(x)$. If we assume that
\begin{align} \label{eq:impl-mut-d-children}
  z_{i+1}^d(x_0) \mid \{ n_i(x) \}_{x \in H_i} \sim \text{Poisson}(\alpha_i \eta_d n_i(x_0)) ,
\end{align}
where $\eta_d$ is the probability for $d$ mutations with $\sum_d \eta_d = 1$, then Equation~\ref{eq:number-of-children} still holds. Naturally, each of the $z_{i+1}^d(x)$ children have to choose their $d$ mutations independently of the others.

To see the analogue between $m_{i+1}(x, x + q)$ and $z_{i+1}(x)$, first let
\begin{align*}
  Q_d = \left \{ q \in Q \Bigm \vert \| q \|_1 = d \right \},
\end{align*}
where $\| \cdot \|_1$ denotes the $L^1$ norm such that $\| q \|_1 = \| (q_1, q_2, \ldots, q_r) \|_1 = \sum_{j=1}^r \vert q_j \vert$. That is, $Q_d$ is the mutation configurations resulting in precisely $d$ mutations. Then
\begin{align*}
  z_{i+1}^d(x) = \sum_{q \in Q_d} m_{i+1}(x, x + q) .
\end{align*}

First the probability of not mutating is treated. Let $\mu_j = \delta_j + \omega_j$ be the mutation rate for the $j$'th locus for $j=1, 2, \ldots, r$ with $\delta_j$ denoting the downwards mutation rate and $\omega_j$ denoting the upwards mutation rate. Then 
\begin{align*}
  \eta_0 = \prod_{j=1}^r (1 - \mu_j ) 
\end{align*}
is the probability of not mutating.

Now the model of choosing the mutating loci is discussed. There are $\binom{r}{d}$ ways to choose the $d$ loci that should mutate. Each of these loci configurations has $2^d$ possible mutation configurations (the size of the cartesian product $\{-1, 1\}^d$). This means that there is a total of $2^d \binom{r}{d}$ possible ways to mutate $d$ times. The probability for mutating to a specific haplotype is determined by the $d$ locus specific upwards and downwards mutation rates.

For mutation category $d$, let
\begin{align*}
  S_d = \left \{ s \subseteq \{ 1, 2, \ldots, r \} \Bigm \vert \vert s \vert = d \right \}
\end{align*}
be a so-called \textit{simple table} with $\binom{r}{d}$ rows. Then the probability that it is exactly the loci $s \in S_d$ that should mutate, is
\begin{align*}
  p(s) = \prod_{j \in s} \mu_j \prod_{j \in s^C} (1-\mu_j) ,
\end{align*}
where $s^C = \{ 1, 2, \ldots, r, \} \setminus s$. Further, the probability of exactly $d$ mutations is
\begin{align*}
  \eta_d = \sum_{s \in S_d} p(s) .
\end{align*}
Hence, Equation~\ref{eq:impl-mut-d-children} is now fully specified. To decide the direction of the mutations, let 
\begin{align*}
  E_d 
  = \left \{ (s, q) \mid s \in S_d, q : s \to \{ -1, 1 \} \right \} 
\end{align*}
be a so-called \textit{extended table} with $2^d \binom{r}{d}$ rows. The function $q$ maps a locus to a mutation direction. Then each row $e = (s, q) \in E_d$ and has probability
\begin{align*}
  p(e) = \prod_{j \in s} p_{j}(q(j)) \prod_{j \in s^C} (1 - p_j(q(j))) ,
\end{align*}
where $p_j(q(j))$ is defined in Equation~\ref{eq:locus-mut-prob}. We still have that the sum of the rows in the extended table is $\eta_d$.

Then for generation $i$, haplotype $x$, and mutation category $d$, we assume that
\begin{align*}
  \{ m_{i+1}(x_0, x_0 + q) \}_{q \in Q_d}
  \mid
  \{ n_i(x) \}_{x \in H_i}
  \sim
  \text{Multinomial} \left ( z_{i+1}^d(x_0) , \left \{ \frac{p(e)}{\sum_{e \in E_d} p(e) } \right \}_{e \in E_d} \right ) .
\end{align*}

Both the simple and extended table for mutation category $d = 1, 2, \ldots, r$ ($d = 0$ does not require this step) are created before the actual simulation starts as the probabilities are constant during the evolution. They are constant because the mutation rates are assumed constant. This is what is done in the \pkg{fwsim} package for all mutation categories, although this may be changed in future releases if the following theoretical limitations turn out to occur in practise, too.

Note that $2^d \binom{r}{d}$, the size of the extended table, is exponentially growing and may become really large for even relatively small $r$ and that the corresponding extended tables take some time to generate. For example, for $r = 16$ and $d=11$ the size of the extended table is $8,945,664$ (the maximal for that choice of $r$), however, it is still possible to be created and used for simulation. Once the tables are created, the simulations run rather smoothly because they are just stored in memory.

On the other hand, the mutation rate would normally be so low that mutations in the categories for even small $d$ may rarely or never happen depending on the population size, which means that these mutation categories are probably better delt with manually as follows. Recall that $\eta_d$ only depends on the simple table, which is small compared to the extended table -- namely a factor of $2^d$ smaller -- and so the simple table can still be calculated to a rather large $r$. When the simple tables are generated, then draw $n$ from $\text{Poisson}(\alpha_{i+1} \eta_d n_i(x))$ and mutate each of the haplotypes manually one at a time by choosing the $d$ loci and their directions randomly according to their probabilities.

\section{Computation time}
\nopagebreak
The simulation method described above is developed with efficiency in mind. To illustrate that efficiency is achieved, the computation time for different parameters have been investigated using a laptop with a 2.40GHz Intel(R) Core(TM) i5 CPU (model M 520). For these computations, \pkg{fwsim} \citep{fwsim025} version 0.2-5 was used.

\begin{figure}[ht]
	\centering
	\includegraphics[width=0.75\textwidth]{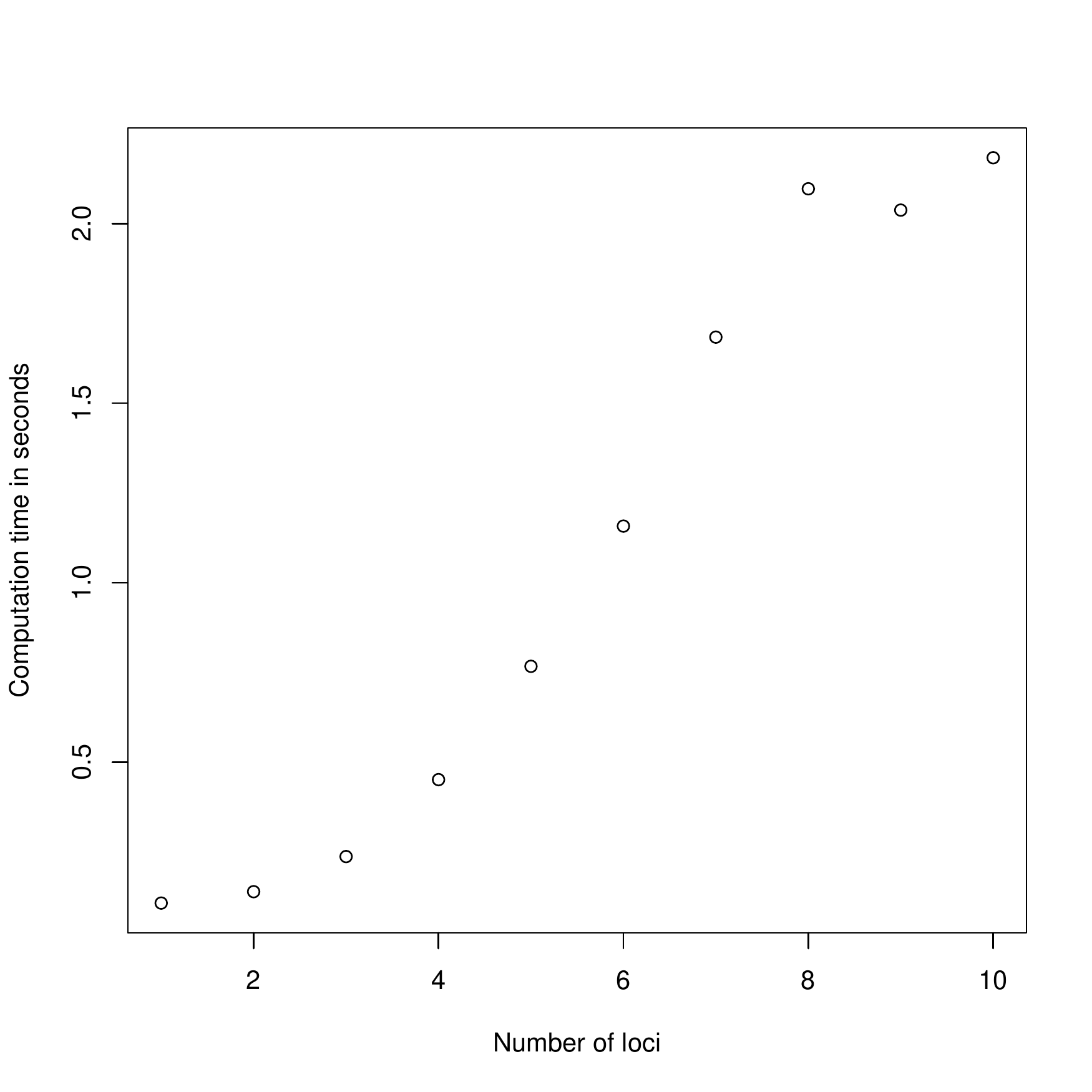}
	\caption{The computation time depending on the number of loci. The initial population size is set to 10,000, the number of generations to 500, the mutation rate to 0.003, the growth parameter to 1 (meaning constant expected population size). The computation time for each number of loci is the median computation time of 10 simulations.}
	\label{fig:time-consumption-loci}
\end{figure}

In Figure~\ref{fig:time-consumption-loci}, the absolute computation time for simulating a population with a varying number of loci is shown. In Figure~\ref{fig:time-consumption-loci}, the computation time for simulating a population with a varying initial population size is shown. Both figures show that the algorithm is quite fast.

\begin{figure}[ht]
	\centering
	\includegraphics[width=0.75\textwidth]{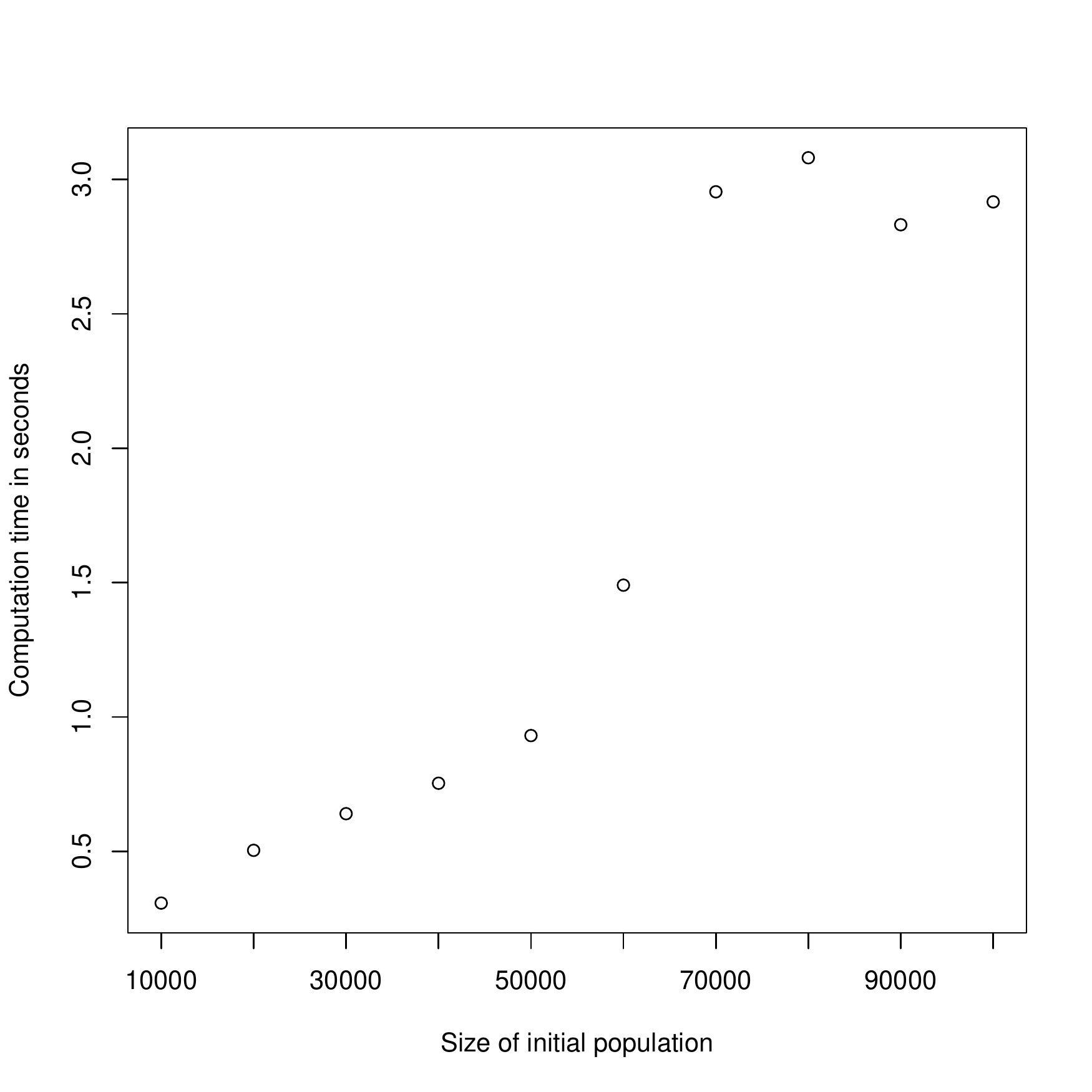}
	\caption{The computation time depending on the initial population size. The number of loci is set to 5, the number of generations to 500, the mutation rate to 0.003, the growth parameter to 1 (meaning constant expected population size). The computation time for each number of loci is the median computation time of 10 simulations.}
	\label{fig:time-consumption-initial-pop}
\end{figure}

In Table~\ref{fig:time-consumption-naive-comparison}, the computation time using \pkg{fwsim} compared to a na\"ive implementation (focusing on individuals rather than haplotypes) of simulating under a Fisher-Wright model is shown. As seen, \pkg{fwsim} is magnitudes faster than a na\"ive implementation: On average, \pkg{fwsim} is almost 2,000 times faster when simulating a population with an initial size of 5,000, no expected growth (by using the growth parameter $\alpha = 1$), and a mutation rate of 0.003 in 100 generations than the na\"ive implementation (focusing on individuals rather than haplotypes). Further, the memory consumption is smaller for \pkg{fwsim} as it uses haplotypes instead of individuals, which means that it is possible to simulate much larger populations than with a na\"ive implementation.

\begin{table}[ht]
  \centering
  \begin{tabular}{rrrrrr}
    $k$ & $g$ & $\mu$ & Speed-up \\\hline
    1,000 & 100 & 0.001 &  145.9 \\
    1,000 & 100 & 0.003 &  127.2 \\
    1,000 & 200 & 0.001 &  307.9 \\
    1,000 & 200 & 0.003 &  372.5 \\
    5,000 & 100 & 0.001 & 2,972.1 \\
    5,000 & 100 & 0.003 & 1,957.0 \\
    5,000 & 200 & 0.001 & 6,848.4 \\
    5,000 & 200 & 0.003 & 4,887.1 \\
    \hline 
  \end{tabular}
  \caption{A comparison of the computation time for \pkg{fwsim} and a na\"ive implementation (focusing on individuals rather than haplotypes). A growth parameter $\alpha = 1$ is used meaning no expected population growth. $k$ is the initial population size, $g$ is the number of generations to evolve, and $\mu$ is the mutation rate. 10 replications for each parameter combination (corresponding to a row in the table) were performed. The speed-up column is the computation time for the na\"ive implementation divided by the computation time for \pkg{fwsim}. This means that \pkg{fwsim} on average is roughly 2,000 times faster to simulate a population with an initial size of 5,000 and a mutation rate of 0.003 in 100 generations than the na\"ive implementation.}
  \label{fig:time-consumption-naive-comparison}
\end{table}

\section{Examples}
\nopagebreak
In this section, some examples are presented. Please refer to \code{?fwsim} in \proglang{R} for more information about usage of the package \pkg{fwsim}. These examples were made using version 0.2-5 of \pkg{fwsim} \citep{fwsim025}.

\subsection{Simple usage}
\nopagebreak

Lauching an \proglang{R} session and typing the code below will show a short example of the model implemented in the package \pkg{fwsim} (\code{k} is the number of individuals in the initial population, \code{g} is the number of generations to evolve, \code{r} number of loci, \code{mu} mutation rate per loci, \code{alpha} is the population size growth rate and \code{trace} is whether to display trace information):
\begin{CodeChunk}
\begin{CodeInput}
library("fwsim")
set.seed(1)
pop <- fwsim(k = 10000, g = 1000, r = 3, mu = 0.003, 
  alpha = 1.001, trace = TRUE)
\end{CodeInput}
\end{CodeChunk}

\begin{figure}[ht]
	\centering
	\includegraphics[width=0.75\textwidth]{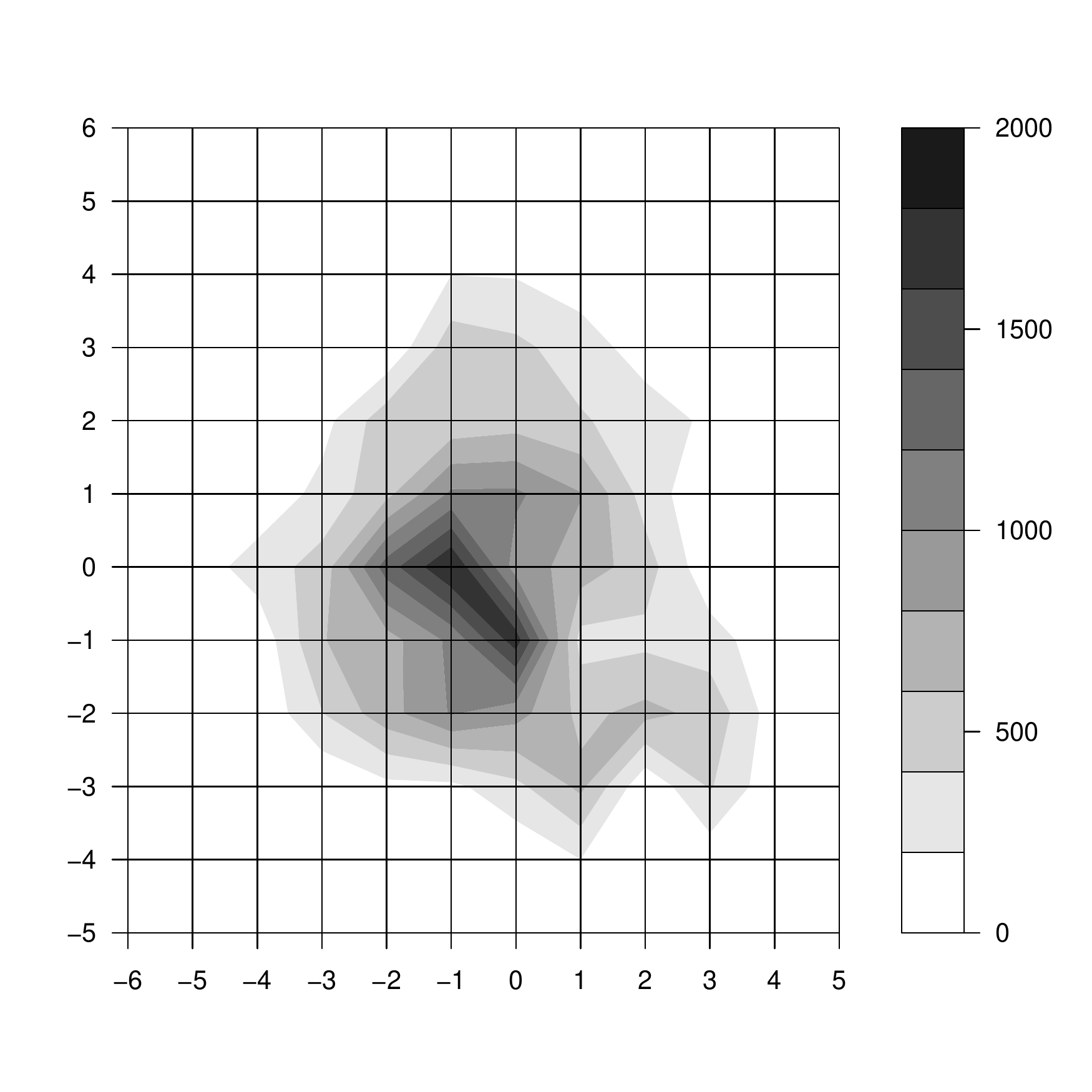}
	\caption{A contour plot of the contingency table of the first two loci. A slight drift from the initial $(0, 0)$ has occured.}
	\label{fig:tab-contour}
\end{figure}

To obtain a contingency table of the first two loci, use the following:
\begin{CodeChunk}
\begin{CodeInput}
sum(pop$haplotypes$N)
\end{CodeInput}
\begin{CodeOutput}
[1] 27672
\end{CodeOutput}
\begin{CodeInput}
xtabs(N ~ Locus1 + Locus2, pop$haplotypes)
\end{CodeInput}
\begin{CodeOutput}
      Locus2
Locus1   -5   -4   -3   -2   -1    0    1    2    3    4    5    6
    -6    0    0    0    0    5    2    0    0    0    0    0    0
    -5    0    0    0   10   85   75   10    0    2    0    0    0
    -4    0    0    0   23   46  301   37  118    1    0    0    0
    -3    0    0   15  394  591  474  266  122  110    5    1    0
    -2    0   11  144  723  717 1302  542  526   17    9   26    0
    -1    0  108  148 1018 1048 1816 1039  453  517  197  138  101
    0     1   30  347  879 1713  901 1038  509  448  184   27   11
    1    34  198  647  552  324  715  810  421  300   90   11    0
    2     0   63   37  659  349  492  314  306  105   10    0    0
    3     0   73  420  540  290   50   30  160    0    0    0    0
    4     0   20   58   94   63    4   41    2    0    0    0    0
    5     0    0    9    0    0    0    0    0    0    0    0    0
\end{CodeOutput}
\end{CodeChunk}
This table is plotted in Figure~\ref{fig:tab-contour}. A slight drift from the initial $(0, 0)$ has occured.

We can also see the $10$ most frequent haplotypes compared to the initial $(0, 0, 0)$ haplotype:
\begin{CodeChunk}
\begin{CodeInput}
pop$haplotypes[order(pop$haplotypes$N, decreasing = TRUE)[1:10], ]
\end{CodeInput}
\begin{CodeOutput}
    Locus1 Locus2 Locus3   N
279     -1      0      0 665
105     -1     -2     -2 539
270     -1      0     -1 517
269     -2      0     -1 509
173      0     -1     -1 482
160      0     -1     -2 423
179      0     -1      0 423
341     -1      1     -1 385
274     -2      0      0 378
241     -1      0     -2 358
\end{CodeOutput}
\begin{CodeInput}
pop$haplotypes[which(apply(apply(pop$haplotypes[, 1:3], 1, abs), 
  2, sum) == 0), ]
\end{CodeInput}
\begin{CodeOutput}
    Locus1 Locus2 Locus3   N
280      0      0      0 255
\end{CodeOutput}
\end{CodeChunk}

\begin{figure}[ht]
	\centering
	\includegraphics[width=0.75\textwidth]{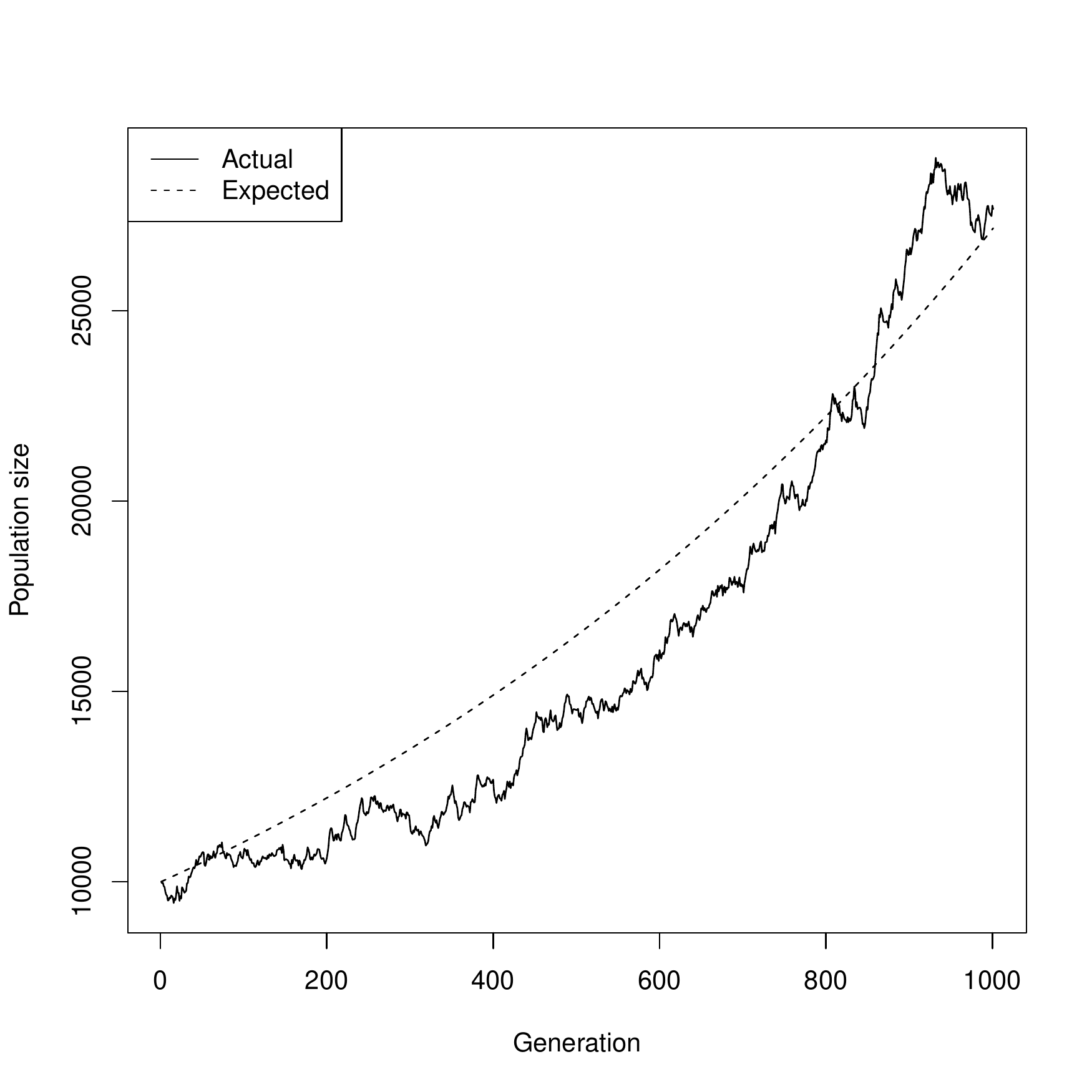}
	\caption{The actual population sizes compared to expected population sizes}
	\label{fig:example-pop-size}
\end{figure}

In Figure~\ref{fig:example-pop-size}, the actual population sizes are compared to expected population sizes. This figure was made with following code:
\begin{CodeChunk}
\begin{CodeInput}
plot(pop$sizes, type = "l", xlab = "Generation", 
  ylab = "Population size", lty = 1)
lines(pop$expected.sizes, lty = 2)
legend("topleft", legend = c("Actual", "Expected"), lty = 1:2)
\end{CodeInput}
\end{CodeChunk}

\subsection{Genetic drift of alleles}
\nopagebreak
To illustrate how genetic drift in terms of changed allele frequencies occurs, the allele frequencies after a different number of generations are recorded. The \pkg{fwsim} package also has the possibility of saving the intermediate populations, which is used to show how allele frequencies change during the evolution. Thus, genetic drift can be investigated as follows (\code{k} is the number of individuals in the initial population, \code{alim} is the limit of which alleles to plot and \code{gs} is which generations to sample allele frequencies from):
\begin{CodeChunk}
\begin{CodeInput}
library("fwsim")
set.seed(1)
alim <- 2
k <- 100000000
g <- 10000
gs <- seq(100, g - 1, by = 100)
pop <- fwsim(g = g, k = k, r = 1, alpha = 1, 
  mu = 0.003, gs = gs, trace = FALSE) 

interhapfreq <- lapply(pop$intermediate.haplotypes[gs], function(hap) {
  tab <- prop.table(xtabs(N ~ Locus1, hap))
  as.vector(tab[which(abs(as.numeric(names(tab))) <= alim)])
})

freq <- data.frame(do.call("rbind", interhapfreq))
colnames(freq) <- (-alim):alim

plot(gs, freq[, alim+1], type = "l", 
  xlab = "Number of generations", 
  ylab = "Frequency", ylim = range(freq))

for (a in 1:alim) {
  i1 <- (alim+1)-a
  i2 <- (alim+1)+a
  lines(gs, freq[, i1], type = "l", lty = a + 1)
  lines(gs, freq[, i2], type = "l", lty = a + 1)
}

others <- 1-apply(freq, 1, sum)
lines(gs, others, type = "l", lty = alim + 2)

legend("topright", 
  legend = c(paste("Allele", c(0, paste("+/-", 1:alim))), 
    "Other alleles"), 
  lty = 1:(alim+2))
\end{CodeInput}
\end{CodeChunk}

\begin{figure}[ht]
	\centering
	\includegraphics[width=0.75\textwidth]{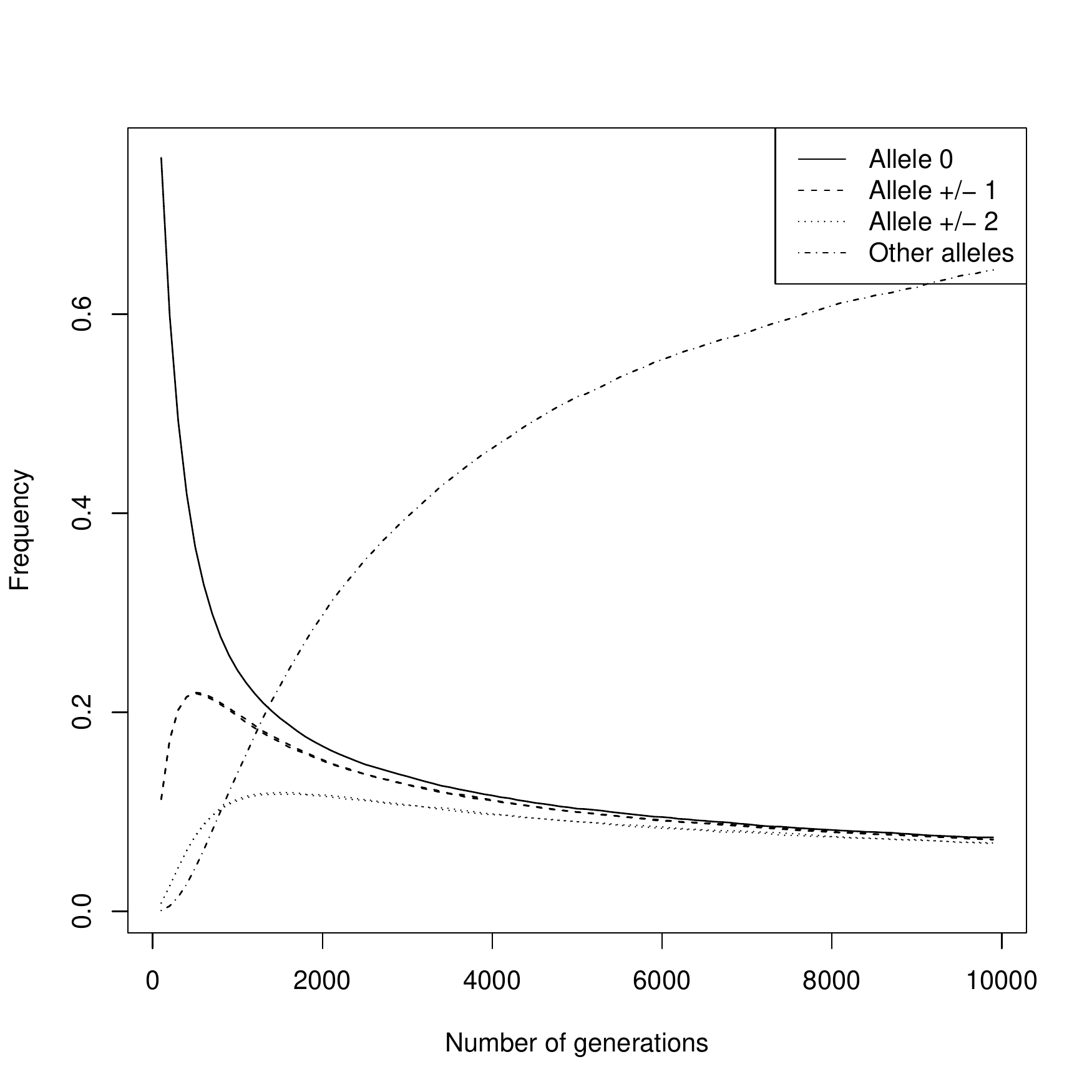}
	\caption{Simulated genetic drift using an initial population of size 100,000,000, a growth of $1$ (meaning no expected growth), and a mutation rate of $0.003$.}
	\label{fig:example-drift}
\end{figure}

Note that we only simulate one locus and set the population size quite large to get the asymptotic behaviour. The resulting plot can be seen in Figure~\ref{fig:example-drift}.

\subsection{Genetic drift of alleles depending on mutation rate}
\nopagebreak
To illustrate how genetic drift in terms of changed allele frequencies for the 0 allele occurs depending on the mutation rate, the allele frequencies after a different number of generations are recorded for populations with different mutation rates. Thus, genetic drift depending on mutation rate may be investigated as follows (\code{k} is the number of individuals in the initial population and \code{gs} is which generations to sample allele frequencies from):
\begin{CodeChunk}
\begin{CodeInput}
library("fwsim")
mus <- c(0.001, 0.002, 0.003)
k <- 100000000
g <- 10000
gs <- seq(100, g - 1, by = 100)

set.seed(1)
freqs <- lapply(mus, function(mu) {
  pop <- fwsim(g = g, k = k, r = 1, alpha = 1, mu = mu, save.gs = gs, 
    trace = FALSE)
  sapply(pop$intermediate.haplotypes[gs], 
    function(hap) hap$N[which(hap[, 1] == 0)] / sum(hap))
})

plot(gs, freqs[[1]], type = "l", 
  xlab = "Number of generations", ylab = "Frequency for allele 0", 
  ylim = range(unlist(lapply(freqs, range))), lty = 1)

for (i in 2:length(mus)) lines(gs, freqs[[i]], type = "l", lty = i)

legend("topright", legend = paste("mu = ", mus, sep = ""), 
  lty = 1:length(mus))
\end{CodeInput}
\end{CodeChunk}

\begin{figure}[ht]
	\centering
	\includegraphics[width=0.75\textwidth]{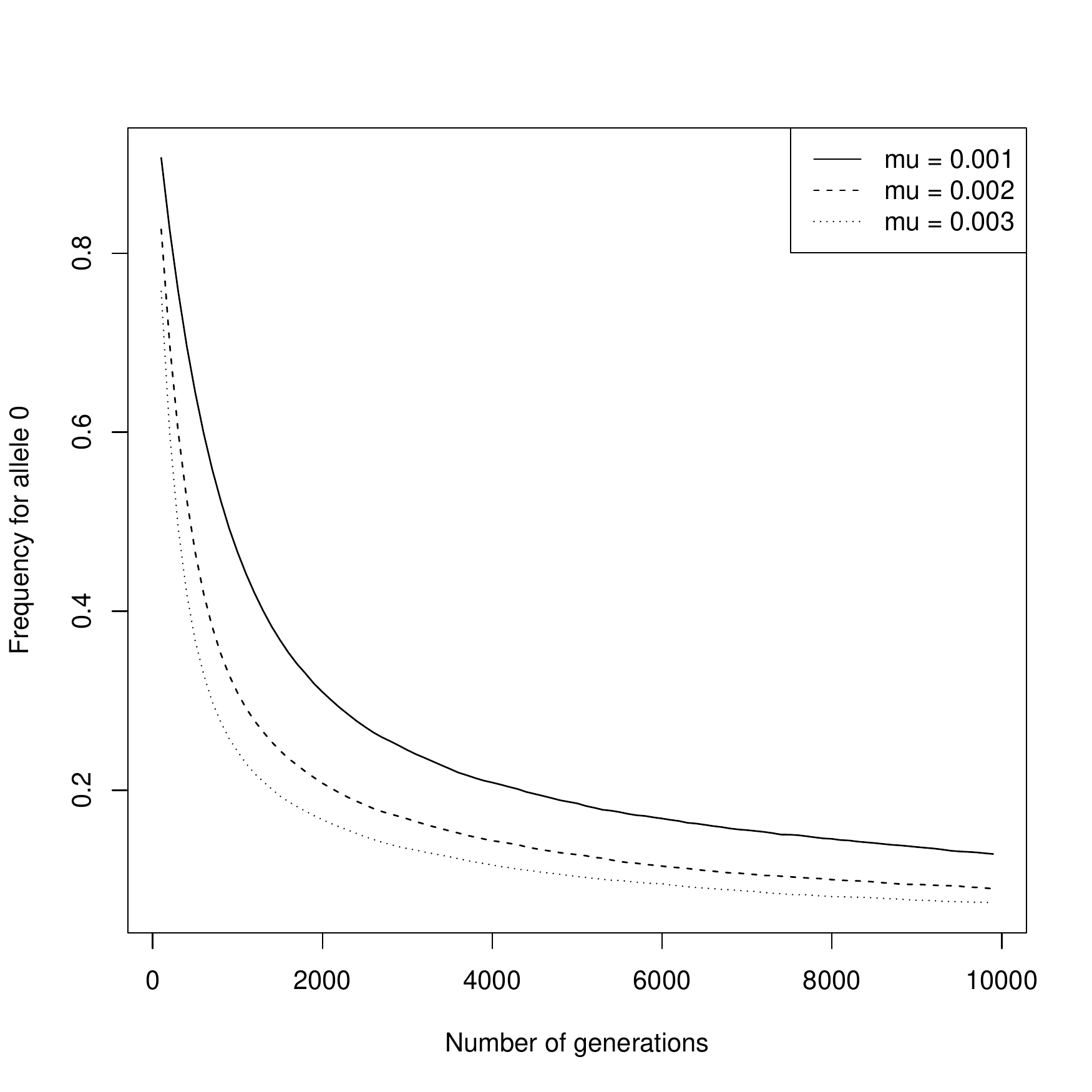}
	\caption{Simulated genetic drift using a population size of 100,000,000 and a growth of $1$ (meaning no expected growth).}
	\label{fig:example-drift-mut}
\end{figure}

Note that we only simulate one locus and set the population size quite large to get the asymptotic behaviour. The resulting plot can be seen in Figure~\ref{fig:example-drift-mut}.

\section*{Acknowledgement}
\nopagebreak
The authors would like to thank Torben Tvedebrink, PhD; Søren Højsgaard, PhD; and Lisbeth Grubbe Nielsen, all Aalborg University, Denmark, for helping us improving the manuscript.

\bibliography{references}

\end{document}